\documentclass[prl,aps,twocolumn,amsmath,amssymb]{revtex4}

\pdfoutput=1
\usepackage{hyperref}
\usepackage{graphicx}

\def\en{\epsilon_{\rm n}}
\def\es{\epsilon_{\rm s}}

\def\np{n_{\parallel}}
\def\ns{n_{\bar{\parallel}}}

\def\beq{\begin{equation}}
\def\eeq{\end{equation}}
\def\bea{\begin{eqnarray}}
\def\eea{\end{eqnarray}}

\def\kt{k_{\rm B}T}

\begin{document}

\title{Random and ordered phases of off-lattice rhombus tiles}

\author{Stephen Whitelam$^1$\footnote{\tt{swhitelam@lbl.gov}}, Isaac Tamblyn$^1$, Peter H. Beton$^2$ and Juan P. Garrahan$^2$} 
\affiliation{$^1$Molecular Foundry, Lawrence Berkeley National Laboratory, 1 Cyclotron Road, Berkeley, CA 94720, USA\\
$^2$School of Physics and Astronomy, University of Nottingham, Nottingham NG7 2RD, UK}

\begin{abstract}

We study the covering of the plane by non-overlapping rhombus tiles, a problem well-studied only in the limiting case of dimer coverings of regular lattices. We go beyond this limit by allowing tiles to take any position and orientation on the plane, to be of irregular shape, and to possess different types of attractive interactions. Using extensive numerical simulations we show that at large tile densities there is a phase transition from a fluid of rhombus tiles to a solid packing with broken rotational symmetry. We observe self-assembly of broken-symmetry phases, even at low densities, in the presence of attractive tile-tile interactions. Depending on tile shape and interactions the solid phase can be random, possessing critical orientational fluctuations, or crystalline. Our results suggest strategies for controlling tiling order in experiments involving `molecular rhombi'.

\end{abstract}

\maketitle

Two-dimensional molecular networks can impart chemical and physical functionalities to semiconductor, metallic and graphite surfaces~\cite{elemans2009molecular,bartels2010tailoring}. They also provide fascinating problems of fundamental science. Small organic molecules, such as p-terphenyl-3,5,3",5"-tetracarboxylic acid (TPTC), can form two-dimensional glassy arrays~\cite{otero2008elementary, blunt2008random} characterized by the absence of long-range translational symmetry. By mapping such arrays onto a rhombus tiling, a classic problem of statistical mechanics~\cite{fisher1961statistical, kasteleyn1963dimer, blšte1982roughening,henley1991random,destainville1998entropy, alet2006classical,papanikolaou2007quantum, castelnovo2007zero,jacobsen2009semiflexible}, one can demonstrate that patterns seen in experiment correspond to configurations of maximum entropy~\cite{blunt2008random}. However, the mapping assumes that the rhombus tiling is constrained by an underlying triangular lattice, and that tiles can effect a complete dimer covering of this lattice, i.e. that TPTC can be represented as what we will call a {\em regular} rhombus whose internal angles are $60^{\circ}$ and $120^{\circ}$. In reality, the graphite lattice used in experiment supports multiple registries and orientations of the molecular overlayer, and molecules do not possess exactly the aspect ratio of the regular rhombus.
\begin{figure}[tr]
\includegraphics[width=\linewidth]{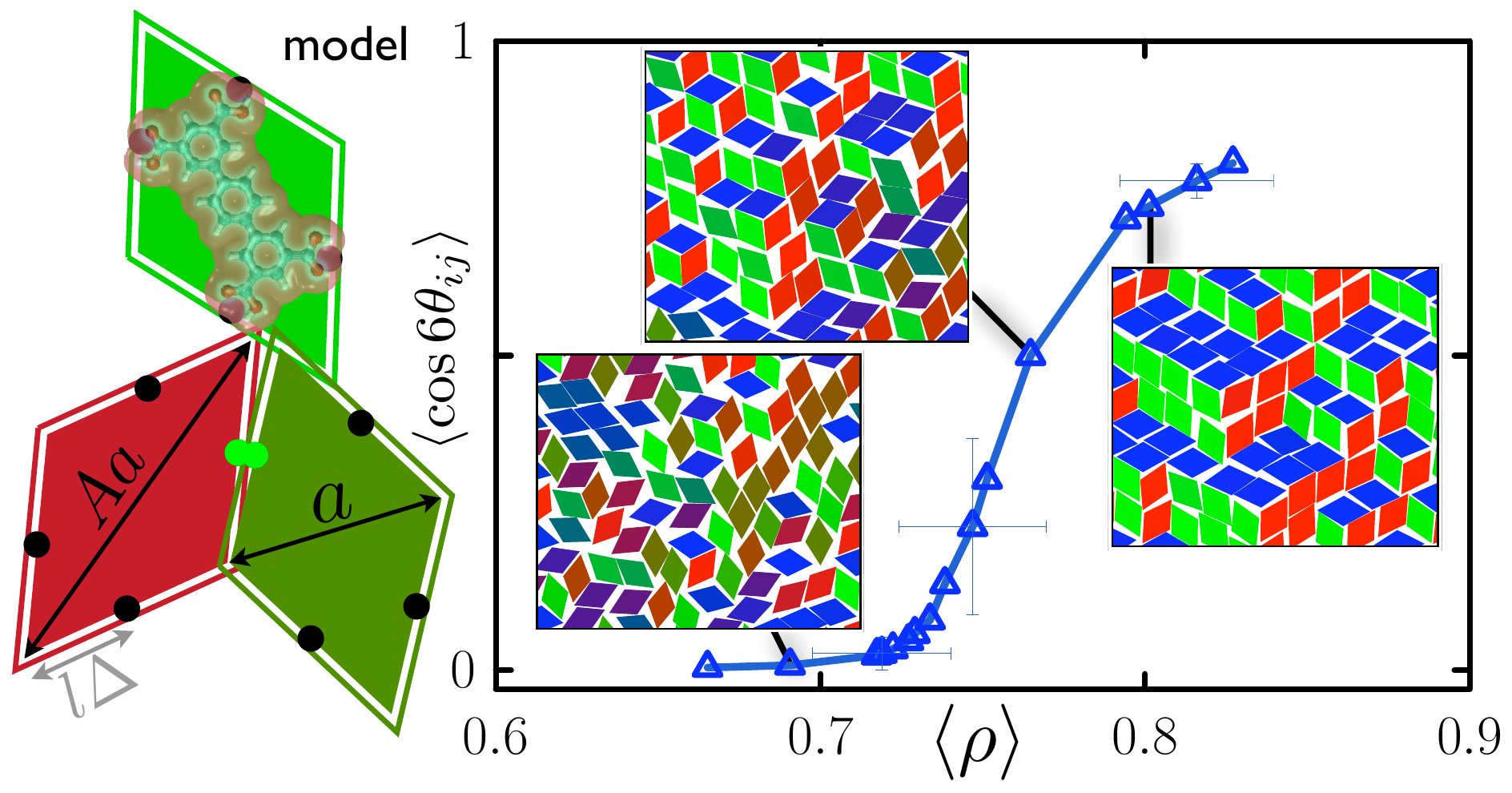}
\caption{\label{fig1} Off-lattice rhombi tile the plane: constant-pressure simulations of hard rhombi of aspect ratio $A=\sqrt{3}$ show the emergence of long range hexatic order at densities $\rho$ above about 70\%. Here and subsequently, rhombi are colored according to their absolute orientations. Left: model geometry, overlaid by isosurface of TPTC ground-state electron density.}
\end{figure}

Motivated by these observations, we use computer simulation to explore the consequences of relaxing the conventional constraints of rhombus tilings: we study rhombi that need not be regular, and that can be placed in any position and with any orientation on the plane. We first show that regular rhombi form hexatic random tilings at high densities, an observation that justifies the conventional lattice approximation. Motivated by recent experiments, we then show that hydrogen bond-like interactions induce the regular rhombus to self-assemble into a random tiling even at low concentrations. We go on to identify geometric and energetic factors that that dictate where tilings are random, ordered, or nonexistent. We argue that these factors might be exploited in experiment to control tiling order.
\begin{figure*}[t!]
\includegraphics[width=\linewidth]{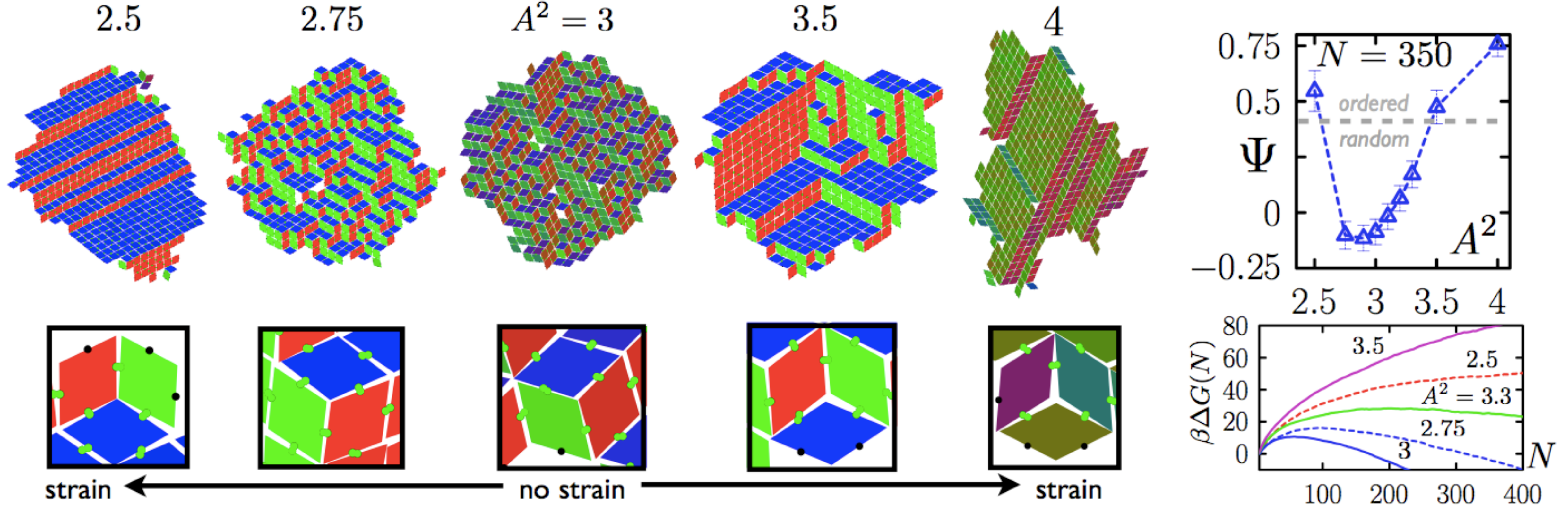}
\caption{\label{fig2} Rhombus tiles with attractive `H-bond' interactions self-assemble into random- and ordered solid phases. Regular rhombi ($A=\sqrt{3}$) assemble as a random solid; sufficiently irregular rhombi form the parallel ordered one (see top right; error bars denote one standard deviation). Snapshot enlargements show that the ordered phase emerges when the 3-particle `boxes' required to form the random tiling become geometrically strained.}
\end{figure*} 

{\em Model.} We simulated the packing and self-assembly of hard rhombi with a long-to-short diagonal aspect ratio $A$ (see Fig.~\ref{fig1}, left) on a featureless two-dimensional substrate; the regular rhombus with internal angles $60^{\circ}$ and $120^{\circ}$ has $A =\sqrt{3}$. The short diagonal length $a$ is typically 1 nm for molecules studied experimentally. Such molecules interact via highly directional hydrogen bonding; to model this we equipped rhombi with patches placed on each edge a distance $l \Delta$ from the small internal angle ($l$ is edge length). Patches on adjacent rhombi give rise to a `specific' energetic reward of $-\es \, \kt$ if they approach closer than a distance $a/10$. To assess the importance of interaction specificity we also considered, in Fig.~\ref{fig4}, a rhombus-shaped nonspecific forcefield of identical aspect ratio and small diagonal length $a_0=11a/10$ (see particle `halo' in Fig.~\ref{fig1}, inset): the overlap of two forcefields triggers a pairwise energetic reward of $-\en \, \kt$. As described in the Supporting Information, we used a collection of custom- and standard Monte Carlo algorithms to study tiling thermodynamics and dynamics. We also used density functional theory (DFT) to locate TPTC within a `design space' of molecular rhombi. We characterized solid order using the parameter $\Psi \equiv (0.608 \np-0.392 \ns)/(0.608 \np+0.392 \ns)$. Here $\np$ is the total number (within the simulation box or the largest cluster, as appropriate) of specific interactions (`H-bonds') between particle pairs whose long diagonals lie closer to being parallel than nonparallel. $\ns$ is the total number of all other H-bonds. This parameter (a generalization of a lattice order parameter defined recently~\cite{tiling-nchem}) allows us to distinguish random tiled structures ($|\Psi| \approx 0$) from crystalline structures with parallel order ($\Psi \lesssim 1$) or nonparallel order ($\Psi \gtrsim -1$; see Fig. S1 for examples of these phases). These distinct solid phases are expected theoretically~\cite{alet2006classical} for on-lattice interacting rhombi, and have been observed experimentally \cite{blunt2008directing}.

{\em Results.} We show in Fig.~\ref{fig1} the results of constant pressure simulations of 768 noninteracting ($\en=\es=0$) regular rhombi. We plot as a function of density the thermal average of the hexatic order parameter $\cos\left(6 \theta_{ij} \right)$, where $\theta_{ij}$ is the angle between vectors pointing along the long diagonals of rhombi $i$ and $j$. The average is taken over all pairs of particles.  Long range hexatic order emerges at densities of about 70\%; such order is consistent with a tiling in which particles point in the three directions of a triangular lattice (simulation box snapshots are shown in Fig. S2).  Notably, the high-density solid phase is not crystalline but random, possessing tiling order $\Psi \approx 0$~\footnote{$\Psi$ is an imperfect (though revealing) measure of order in systems without H-bond interactions. We verified that $\Psi \approx 0$ for (phantom) symmetric patches and for phantom patches placed under the peak of the curve in Fig. S3.}. It has a high degree of orientational order but no translational order. The random phase is known from theoretical studies to possess critical fluctuations in tile orientation characteristic of entropically-stabilized `Coulomb phase' systems~\cite{henley1991random}. Random tilings are seen in the standard limiting case of regular rhombi constrained to an underlying triangular lattice; the spontaneous emergence of similar order here, driven only by rhombus shape, justifies the approximations inherent in that limiting case.
\begin{figure*}[ht]
\includegraphics[width=\linewidth]{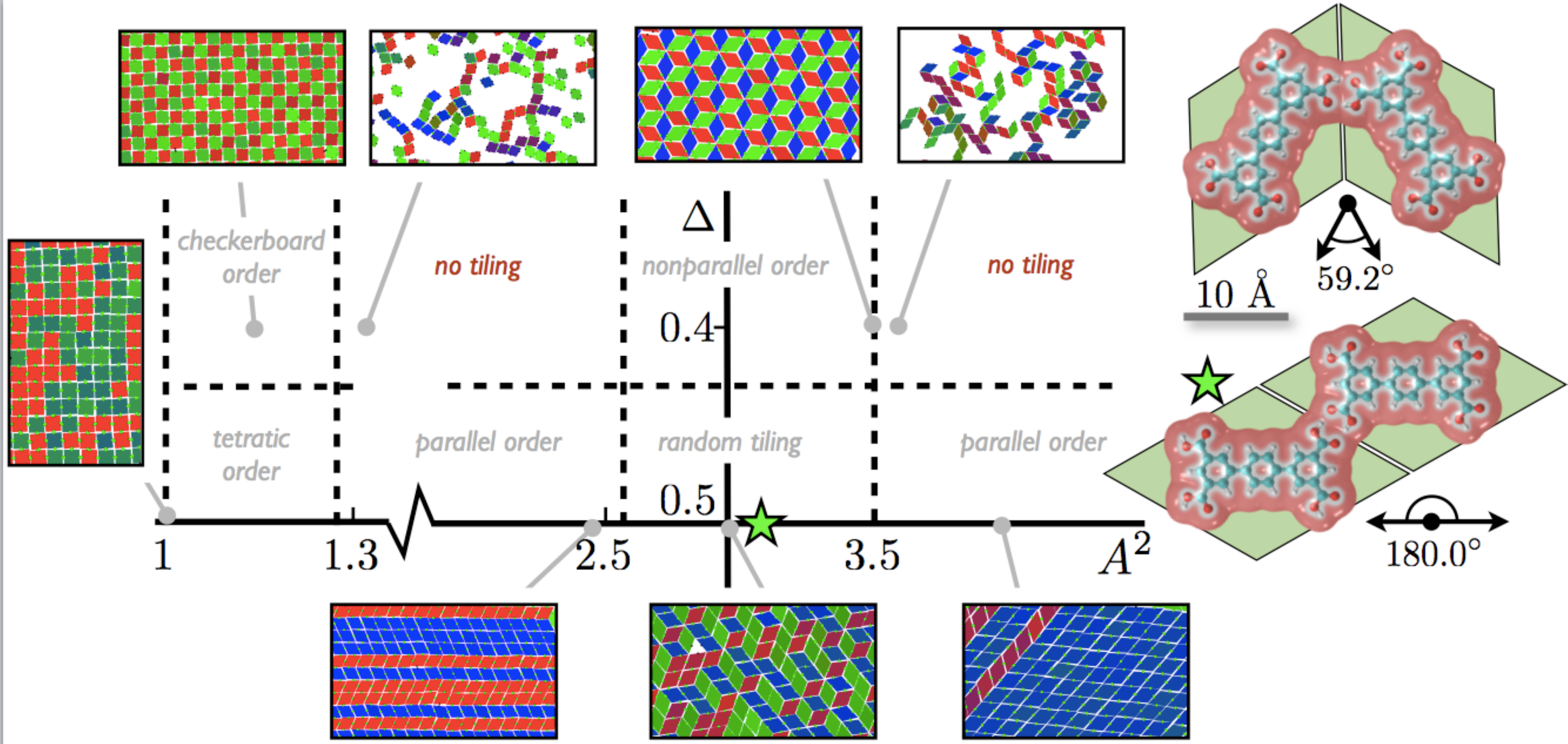}
\caption{\label{fig3} Design space of `molecular rhombi'. Schematic phase diagram of aspect ratio $A$ and patch placement $\Delta$ (phase boundaries are approximate). DFT calculations reveal that TPTC (right) corresponds to a nearly-regular rhombus whose patches are placed near-centrally.}
\end{figure*}

We next verified that equipping rhombi with symmetric ($\Delta=1/2$) `H-bond' interactions allows them to self-assemble, even at low densities, into a random tiling. In Fig.~\ref{fig2} we show results from simulations in which interacting rhombi ($\es=5.2$) were allowed to deposit on a substrate, with chemical potential chosen so that a dense cluster of regular rhombi faced a free energy barrier of about 10 $\kt$ to its nucleation. Nuclei were grown using umbrella sampling.  Nuclei of regular rhombi have a value of $\Psi$ slightly less than zero, showing them to be random~\cite{stannard2010entropically} but with a slight bias for the nonparallel binding mode. This bias results from the fact that for rhombi in an orientationally-ordered tiling, the inter-patch distance between parallel neighbors exceeds, slightly, the inter-patch distance between nonparallel neighbors. This bias can be annulled by shifting patches slightly towards the small internal angle (by contrast, a {\em large} shift exacerbates this bias and drives the emergence of nonparallel order; see Fig. S3). Notably, assemblies display the topologically interesting triangular defects seen in real- and on-lattice networks~\cite{blunt2008random} (a resulting tiling is shown in Fig. S4). 

However, perturbing the aspect ratio of rhombi away from the regular value of $A=\sqrt{3}$ impairs their ability to form a random tiling. In Fig.~\ref{fig2} we show that near-regular rhombi still form clusters having values of $\Psi$ close to zero, an observation that validates the study of TPTC on the lattice~\cite{blunt2008random}. But sufficiently irregular rhombi favor the parallel ordered phase, characterized by large positive $\Psi$. The geometrical reason for the emergence of this phase is shown in the bottom panel of Fig.~\ref{fig2}: the random phase contains 3-particle `boxes' of rhombi that knit together domains of parallel tiles. Boxes form readily when the large internal angle of the rhombus is $120^{\circ}$, but departures from this angle strain boxes and eventually suppress their formation, driving the emergence of a parallel ordered phase. Notably, analog molecular rhombi with larger aspect ratios than TPTC have recently been shown to form biased random tilings with $\Psi \approx 0.2$~\cite{tiling-nchem}; our calculations suggest that geometric factors can account for the symmetry breaking leading to such phase behavior. Suppressing box formation also increases the work of formation (bottom right) of a cluster of $N$ tiles.

The design space of `molecular rhombi' therefore admits ordered and random tiled phases, as summarized in Fig.~\ref{fig3}. TPTC is well `designed' as a random-tiling agent. Our DFT calculations show binding modes that are parallel and 59.2$^{\circ}$ disposed, consistent with a rhombus aspect ratio $A \approx \sqrt{3.1}$, and a patch center $\Delta$ for these motifs of 0.50. Both parameters lie well within the limits of the random phase. More generally, we predict that tiling order near the square-tile limit $A = 1$ is tetratic rather than hexatic, and we predict the existence of a `checkerboard' phase. We also predict the breakdown of order: tiles cannot form dense assemblies when tile shape favors parallel binding, but patch placement cannot accommodate it. This conflict defines the geometric limits within which molecular rhombi tile, with the proviso that the precise location of these limits will depend on the ratio of molecular size and the range of H-bond interactions, which we have represented only roughly. 
\begin{figure*}
\includegraphics[width=\linewidth]{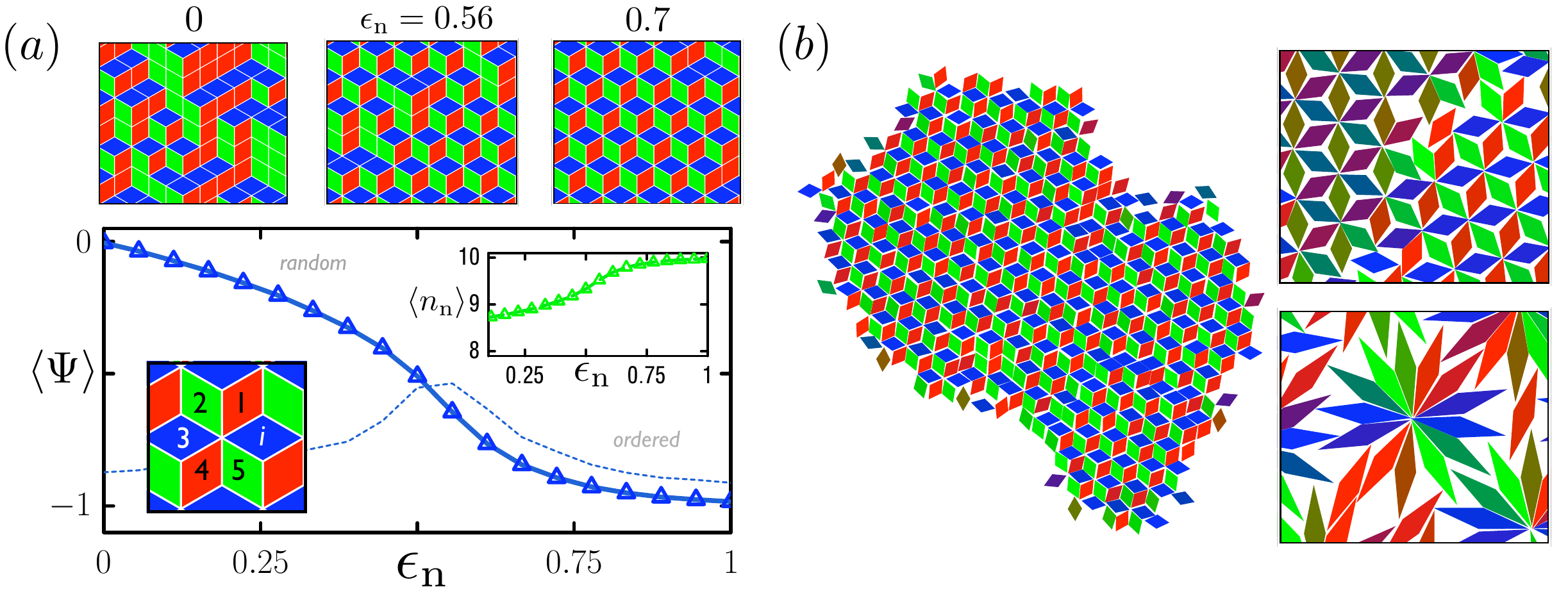}
\caption{\label{fig4} Interactions other than H-bond attractions may act to select tiling phases. (a) We consider regular rhombi interacting nonspecifically, initially arranged as a random tiling of a triangular lattice. We allow sampling of dense phases using the 3-particle rotation algorithm (see SI); we plot the thermal average of the tiling order parameter $\Psi$ as a function of $\en$, the strength of the nonspecific attraction (peaked line is shifted and scaled variance of $\Psi$). This attraction favors high-order vertices of the kind found readily in the nonparallel ordered tiling (inset bottom left: particle $i$ has two lots of five vertex neighbors), and if strong enough it destabilizes the random tiling. Inset top right: average number $n_{\rm n}$ of pairwise nonspecific interactions made by particles. (b) Self-assembly of regular and irregular rhombi driven by the nonspecific interaction. We show configurations taken from dynamical trajectories of rhombi equipped with nonspecific attractions of fixed strength, for the cases $A^2=3,5,20$ (clockwise from left). The nonspecific attraction favors the formation of tiling vertices, which for the regular rhombus leads to the emergence of the nonparallel ordered tiling. However, as $A$ departs enough from $\sqrt{3}$ the nonparallel ordered tiling cannot exist. Open structures result.}
\end{figure*}

Real molecules also possess interactions, such as van der Waals- or solvent-mediated forces, that are less specific than hydrogen bonding. We have not represented such effects, but we can demonstrate the potential importance of nonspecific forces. In Fig.~\ref{fig4} we show that a short range rhombus-shaped attraction effects a bias in favor of a nonparallel ordered tiling. Through this nonspecific interaction, rhombi `see' energetically their neighbors at each vertex in a tiling. In tilings with perfect nonparallel order each rhombus has ten such neighbors, five at each corner of its long diagonal (inset bottom left). In a random tiling, rhombi have fewer vertex neighbors (inset top right), and a strong enough nonspecific interaction induces a phase transition from the random phase to the ordered one. In off-lattice simulations such an attraction can induce a nonparallel bias in tilings of regular rhombi (with or without H-bond interactions), and induce irregular rhombi to form patterns unlike any of the known dense rhombus tilings (Fig.~\ref{fig4}b). Recent experiments suggest that forces {\em other} than molecules' H-bonds can dictate tiling order: TPTC forms random tilings with a pronounced nonparallel bias ($-0.5 \lesssim \Psi \lesssim -0.1$) in certain solvents~\cite{tiling-nchem}. While our nonspecific interaction is not a realistic representation of intermolecular forces, its effect illustrates a possible mechanism for the establishment of a similar bias.

We have explored rhombus tilings beyond the standard limit of dimer coverings of the regular lattice, identifying physical mechanisms that lead to the emergence of ordered and random phases. We have also made predictions for novel behaviors that might be realized using `molecular rhombi'. The methods described here can be used to study the packing and self-assembly of a wide variety of tiles, aiding the search for molecular networks that possess interesting properties such as quasicrystalline order~\cite{divincenzo1999quasicrystals}.

{\em Acknowledgements} We thank Lester Hedges for assistance with histogram reweighting, and Baron Peters for discussions. This work was done as part of a User project at the Molecular Foundry, Lawrence Berkeley National Laboratory, and was supported by the Director, Office of Science, Office of Basic Energy Sciences, of the U.S. Department of Energy under Contract No. DE-AC02--05CH11231. IT acknowledges support from NSERC.


\begin{thebibliography}{18}
\expandafter\ifx\csname natexlab\endcsname\relax\def\natexlab#1{#1}\fi
\expandafter\ifx\csname bibnamefont\endcsname\relax
  \def\bibnamefont#1{#1}\fi
\expandafter\ifx\csname bibfnamefont\endcsname\relax
  \def\bibfnamefont#1{#1}\fi
\expandafter\ifx\csname citenamefont\endcsname\relax
  \def\citenamefont#1{#1}\fi
\expandafter\ifx\csname url\endcsname\relax
  \def\url#1{\texttt{#1}}\fi
\expandafter\ifx\csname urlprefix\endcsname\relax\def\urlprefix{URL }\fi
\providecommand{\bibinfo}[2]{#2}
\providecommand{\eprint}[2][]{\url{#2}}

\bibitem[{\citenamefont{Elemans et~al.}(2009)\citenamefont{Elemans, Lei, and
  De~Feyter}}]{elemans2009molecular}
\bibinfo{author}{\bibfnamefont{J.}~\bibnamefont{Elemans}},
  \bibinfo{author}{\bibfnamefont{S.}~\bibnamefont{Lei}}, \bibnamefont{and}
  \bibinfo{author}{\bibfnamefont{S.}~\bibnamefont{De~Feyter}},
  \bibinfo{journal}{Angewandte Chemie International Edition}
  \textbf{\bibinfo{volume}{48}}, \bibinfo{pages}{7298} (\bibinfo{year}{2009}).

\bibitem[{\citenamefont{Bartels}(2010)}]{bartels2010tailoring}
\bibinfo{author}{\bibfnamefont{L.}~\bibnamefont{Bartels}},
  \bibinfo{journal}{Nature Chemistry} \textbf{\bibinfo{volume}{2}},
  \bibinfo{pages}{87} (\bibinfo{year}{2010}).

\bibitem[{\citenamefont{Otero et~al.}(2008)\citenamefont{Otero, Lukas, Kelly,
  Xu, L{\ae}gsgaard, Stensgaard, Kantorovich, and
  Besenbacher}}]{otero2008elementary}
\bibinfo{author}{\bibfnamefont{R.}~\bibnamefont{Otero}},
  \bibinfo{author}{\bibfnamefont{M.}~\bibnamefont{Lukas}},
  \bibinfo{author}{\bibfnamefont{R.}~\bibnamefont{Kelly}},
  \bibinfo{author}{\bibfnamefont{W.}~\bibnamefont{Xu}},
  \bibinfo{author}{\bibfnamefont{E.}~\bibnamefont{L{\ae}gsgaard}},
  \bibinfo{author}{\bibfnamefont{I.}~\bibnamefont{Stensgaard}},
  \bibinfo{author}{\bibfnamefont{L.}~\bibnamefont{Kantorovich}},
  \bibnamefont{and}
  \bibinfo{author}{\bibfnamefont{F.}~\bibnamefont{Besenbacher}},
  \bibinfo{journal}{Science} \textbf{\bibinfo{volume}{319}},
  \bibinfo{pages}{312} (\bibinfo{year}{2008}).

\bibitem[{\citenamefont{Blunt et~al.}(2008{\natexlab{a}})\citenamefont{Blunt,
  Russell, Gim{\'e}nez-L{\'o}pez, Garrahan, Lin, Schr{\"o}der, Champness, and
  Beton}}]{blunt2008random}
\bibinfo{author}{\bibfnamefont{M.}~\bibnamefont{Blunt}},
  \bibinfo{author}{\bibfnamefont{J.}~\bibnamefont{Russell}},
  \bibinfo{author}{\bibfnamefont{M.}~\bibnamefont{Gim{\'e}nez-L{\'o}pez}},
  \bibinfo{author}{\bibfnamefont{J.}~\bibnamefont{Garrahan}},
  \bibinfo{author}{\bibfnamefont{X.}~\bibnamefont{Lin}},
  \bibinfo{author}{\bibfnamefont{M.}~\bibnamefont{Schr{\"o}der}},
  \bibinfo{author}{\bibfnamefont{N.}~\bibnamefont{Champness}},
  \bibnamefont{and} \bibinfo{author}{\bibfnamefont{P.}~\bibnamefont{Beton}},
  \bibinfo{journal}{Science} \textbf{\bibinfo{volume}{322}},
  \bibinfo{pages}{1077} (\bibinfo{year}{2008}{\natexlab{a}}).

\bibitem[{\citenamefont{Fisher}(1961)}]{fisher1961statistical}
\bibinfo{author}{\bibfnamefont{M.}~\bibnamefont{Fisher}},
  \bibinfo{journal}{Physical Review} \textbf{\bibinfo{volume}{124}},
  \bibinfo{pages}{1664} (\bibinfo{year}{1961}).

\bibitem[{\citenamefont{Kasteleyn}(1963)}]{kasteleyn1963dimer}
\bibinfo{author}{\bibfnamefont{P.}~\bibnamefont{Kasteleyn}},
  \bibinfo{journal}{Journal of Mathematical Physics}
  \textbf{\bibinfo{volume}{4}}, \bibinfo{pages}{287} (\bibinfo{year}{1963}).

\bibitem[{\citenamefont{Bl{\"o}te and Hilhorst}(1982)}]{blšte1982roughening}
\bibinfo{author}{\bibfnamefont{H.}~\bibnamefont{Bl{\"o}te}} \bibnamefont{and}
  \bibinfo{author}{\bibfnamefont{H.}~\bibnamefont{Hilhorst}},
  \bibinfo{journal}{J. Phys. A} \textbf{\bibinfo{volume}{15}},
  \bibinfo{pages}{L631} (\bibinfo{year}{1982}).

\bibitem[{\citenamefont{Henley}(1991)}]{henley1991random}
\bibinfo{author}{\bibfnamefont{C.}~\bibnamefont{Henley}},
  \bibinfo{journal}{Quasicrystals: the state of the art} pp.
  \bibinfo{pages}{429--524} (\bibinfo{year}{1991}).

\bibitem[{\citenamefont{Cohn et~al.}(2001)\citenamefont{Cohn, Kenyon, and
  Propp}}]{destainville1998entropy}
\bibinfo{author}{\bibfnamefont{H.}~\bibnamefont{Cohn}},
  \bibinfo{author}{\bibfnamefont{R.}~\bibnamefont{Kenyon}}, \bibnamefont{and}
  \bibinfo{author}{\bibfnamefont{J.}~\bibnamefont{Propp}},
  \bibinfo{journal}{Journal of the American Mathematical Society}
  \textbf{\bibinfo{volume}{14}}, \bibinfo{pages}{297} (\bibinfo{year}{2001}).

\bibitem[{\citenamefont{Alet et~al.}(2006)\citenamefont{Alet, Ikhlef, Jacobsen,
  Misguich, and Pasquier}}]{alet2006classical}
\bibinfo{author}{\bibfnamefont{F.}~\bibnamefont{Alet}},
  \bibinfo{author}{\bibfnamefont{Y.}~\bibnamefont{Ikhlef}},
  \bibinfo{author}{\bibfnamefont{J.}~\bibnamefont{Jacobsen}},
  \bibinfo{author}{\bibfnamefont{G.}~\bibnamefont{Misguich}}, \bibnamefont{and}
  \bibinfo{author}{\bibfnamefont{V.}~\bibnamefont{Pasquier}},
  \bibinfo{journal}{Physical Review E} \textbf{\bibinfo{volume}{74}},
  \bibinfo{pages}{041124} (\bibinfo{year}{2006}).

\bibitem[{\citenamefont{Papanikolaou et~al.}(2007)\citenamefont{Papanikolaou,
  Luijten, and Fradkin}}]{papanikolaou2007quantum}
\bibinfo{author}{\bibfnamefont{S.}~\bibnamefont{Papanikolaou}},
  \bibinfo{author}{\bibfnamefont{E.}~\bibnamefont{Luijten}}, \bibnamefont{and}
  \bibinfo{author}{\bibfnamefont{E.}~\bibnamefont{Fradkin}},
  \bibinfo{journal}{Physical Review B} \textbf{\bibinfo{volume}{76}},
  \bibinfo{pages}{134514} (\bibinfo{year}{2007}).

\bibitem[{\citenamefont{Castelnovo et~al.}(2007)\citenamefont{Castelnovo,
  Chamon, Mudry, and Pujol}}]{castelnovo2007zero}
\bibinfo{author}{\bibfnamefont{C.}~\bibnamefont{Castelnovo}},
  \bibinfo{author}{\bibfnamefont{C.}~\bibnamefont{Chamon}},
  \bibinfo{author}{\bibfnamefont{C.}~\bibnamefont{Mudry}}, \bibnamefont{and}
  \bibinfo{author}{\bibfnamefont{P.}~\bibnamefont{Pujol}},
  \bibinfo{journal}{Annals of Physics} \textbf{\bibinfo{volume}{322}},
  \bibinfo{pages}{903} (\bibinfo{year}{2007}).

\bibitem[{\citenamefont{Jacobsen and Alet}(2009)}]{jacobsen2009semiflexible}
\bibinfo{author}{\bibfnamefont{J.}~\bibnamefont{Jacobsen}} \bibnamefont{and}
  \bibinfo{author}{\bibfnamefont{F.}~\bibnamefont{Alet}},
  \bibinfo{journal}{Physical Review Letters} \textbf{\bibinfo{volume}{102}},
  \bibinfo{pages}{145702} (\bibinfo{year}{2009}).

\bibitem[{\citenamefont{Stannard et~al.}(2011)}]{tiling-nchem}
\bibinfo{author}{\bibfnamefont{A.}~\bibnamefont{Stannard}}
  \bibnamefont{et~al.}, \bibinfo{journal}{Nat. Chem., in press}
  (\bibinfo{year}{2011}).

\bibitem[{\citenamefont{Blunt et~al.}(2008{\natexlab{b}})\citenamefont{Blunt,
  Lin, del Carmen Gimenez-Lopez, Schr{\"o}der, Champness, and
  Beton}}]{blunt2008directing}
\bibinfo{author}{\bibfnamefont{M.}~\bibnamefont{Blunt}},
  \bibinfo{author}{\bibfnamefont{X.}~\bibnamefont{Lin}},
  \bibinfo{author}{\bibfnamefont{M.}~\bibnamefont{del Carmen Gimenez-Lopez}},
  \bibinfo{author}{\bibfnamefont{M.}~\bibnamefont{Schr{\"o}der}},
  \bibinfo{author}{\bibfnamefont{N.}~\bibnamefont{Champness}},
  \bibnamefont{and} \bibinfo{author}{\bibfnamefont{P.}~\bibnamefont{Beton}},
  \bibinfo{journal}{Chem. Commun.} pp. \bibinfo{pages}{2304--2306}
  (\bibinfo{year}{2008}{\natexlab{b}}).

\bibitem[{\citenamefont{Stannard et~al.}(2010)\citenamefont{Stannard, Blunt,
  Beton, and Garrahan}}]{stannard2010entropically}
\bibinfo{author}{\bibfnamefont{A.}~\bibnamefont{Stannard}},
  \bibinfo{author}{\bibfnamefont{M.}~\bibnamefont{Blunt}},
  \bibinfo{author}{\bibfnamefont{P.}~\bibnamefont{Beton}}, \bibnamefont{and}
  \bibinfo{author}{\bibfnamefont{J.}~\bibnamefont{Garrahan}},
  \bibinfo{journal}{Physical Review E} \textbf{\bibinfo{volume}{82}},
  \bibinfo{pages}{041109} (\bibinfo{year}{2010}).

\bibitem[{\citenamefont{DiVincenzo and
  Steinhardt}(1999)}]{divincenzo1999quasicrystals}
\bibinfo{author}{\bibfnamefont{D.}~\bibnamefont{DiVincenzo}} \bibnamefont{and}
  \bibinfo{author}{\bibfnamefont{P.}~\bibnamefont{Steinhardt}},
  \emph{\bibinfo{title}{{Quasicrystals: the State of the Art}}}
  (\bibinfo{publisher}{World Scientific Pub Co Inc}, \bibinfo{year}{1999}),
  ISBN \bibinfo{isbn}{9810241569}.

\end{thebibliography}

\end{document}